\documentclass[multphys,vecphys]{promult}

\usepackage{makeidx}     % allows index generation
\usepackage{graphicx}    % standard LaTeX graphics tool
\usepackage{multicol}    % used for the two-column index
\makeindex             % used for the subject index
\newcommand{\cmd}[1]{\texttt{#1}}

\begin{document}

\title*{The software for the robotization of the TROBAR telescope}
\author{Mauro Stefanon\inst{1}}
\institute{Observatori Astronomic  Universitat de Val\'encia - Edifici Instituts d'Investigaci\'o Pol\'igon La Coma - 46980 Paterna, Val\'encia - Spain}
%\and Department of Terrestrian Astronomy, University of Antartica, South Pole 9999, Antartica} 
%
% 
%
\maketitle

The Telescopi ROBotic de ARas (TROBAR) is a new robotic facility built at Aras de Los Olmos (Valencia-Spain). 
This is a 60cm  telescope equipped with a 4kx4k optical camera, corresponding to 30x30 arcmin$^2$ FoV, and it will be primarily used for a systematic search of Ha emitting stars in the Galactic Plane to a depth of $\approx$14mag. Both data acquisition and reduction will be performed automatically. 
The robotization of data acquisition is now entering its final coding phase while the development of the data reduction pipeline has just started.

\section{Introduction}
\label{sec:1}
% Always give a unique label
% and use \ref{<label>} for cross-references
% and \cite{<label>} for bibliographic references
% use \sectionmark{}
% to alter or adjust the section heading in the running head
Galactic H$\alpha$ emitting objects are tracers of pre and post main-sequence stars as well as of nebulae, cataclysmic variables, Be stars and other more exotic objects like Luminous Blue Variables (LBV) and Wolf-Rayet stars. IPHAS (\cite{iphas1}, \cite{iphas2}, \cite{iphas3}, \cite{iphas4}) the most complete survey of the galactic plane carried out so far, is complete in the magnitude range $r'=13$ to $20$ for $-5<|b|<5$. The classical surveys, mostly based on objective-prism photographic observations, are complete up to magnitude 9 (see for example \cite{old1} and \cite{old2}). In this context, the main aim of our project is to carry out a photometric survey covering the existing gap down to $r'\approx$14 with observations and data reduction automatically performed.

\section{Telescope and location}
\label{sec:telescope}
TROBAR is located at the Observatori Astronomic de Aras (OAA), approximately 100Km  north-west of Valencia, at an altitude of 1330m, in a region of low light pollution. %(see Figure \ref{fig:site}). 

The telescope, realized by Teleskoptechnik Halfmann, has a main mirror of 60 cm in diameter, with a classical Ritchey-Chr\'etien optical scheme; a Nasmyth focus is present, to which an optical camera is attached. The telescope can slew as fast as 10deg/sec allowing to point any region of the sky in less than a minute. A filterwheel hosts standard Sloan u, g, r, i, z,  Str\"{o}mgren u, b, y, v plus two $H{\alpha}$ filters (one narrow and one medium-band).
Table \ref{tab:main_features} lists the main features of the telescope.
\begin{table}[tb]
\centering
\caption{TROBAR main features}
\label{tab:1}       % Give a unique label
\begin{tabular}{lc}
\hline\noalign{\smallskip}
Device  & Value  \\
\noalign{\smallskip}\hline\noalign{\smallskip}
M1 diameter  & 60 cm \\
Focal length & 4800 mm \\
Mount & Alt-az - Nasmyh focus \\
Camera & Fairchild $4K \times 4K$ optical camera \\
FoV & $30' \times 30'$ \\
Filters & Sloan u, g, r, i, z; Str\"{o}mgren u, b, v, y; $H\alpha$ (narrow+medium band) \\
\noalign{\smallskip}\hline
\end{tabular}
\label{tab:main_features}
\end{table}

Low level control of telescope pointing capabilities is done by the \emph{Pilar} software provided by $4\Pi$ Systems Gmb company, which supplies a socket interface to the TCS commands.

The optical detector is a Fairchild Peregrine 486 back-illuminated CCD, providing an array of $4096\times 4096$ pixels of $15 \mu m \times 15 \mu m$ read by four amplifiers; the  readout noise is $4e^-$.

The telescope can be both remotely controlled and robotically operated and a set of commands also allows to perform almost all the operations via scripting procedure. Here we will outline its robotic capabilities.

\section{System description}
The operations which can be performed by the telescope comprise the acquisition of calibration frames (bias, darks, sky flat field and focus) and the observation of scientific targets.

All the software is written in python, with the Object Oriented paradigm. All astronomical calculations are performed through the pyephem libraries\footnote{http://rhodesmill.org/pyephem/}.

\begin{figure}[tb]
\centering
\includegraphics[height=12cm]{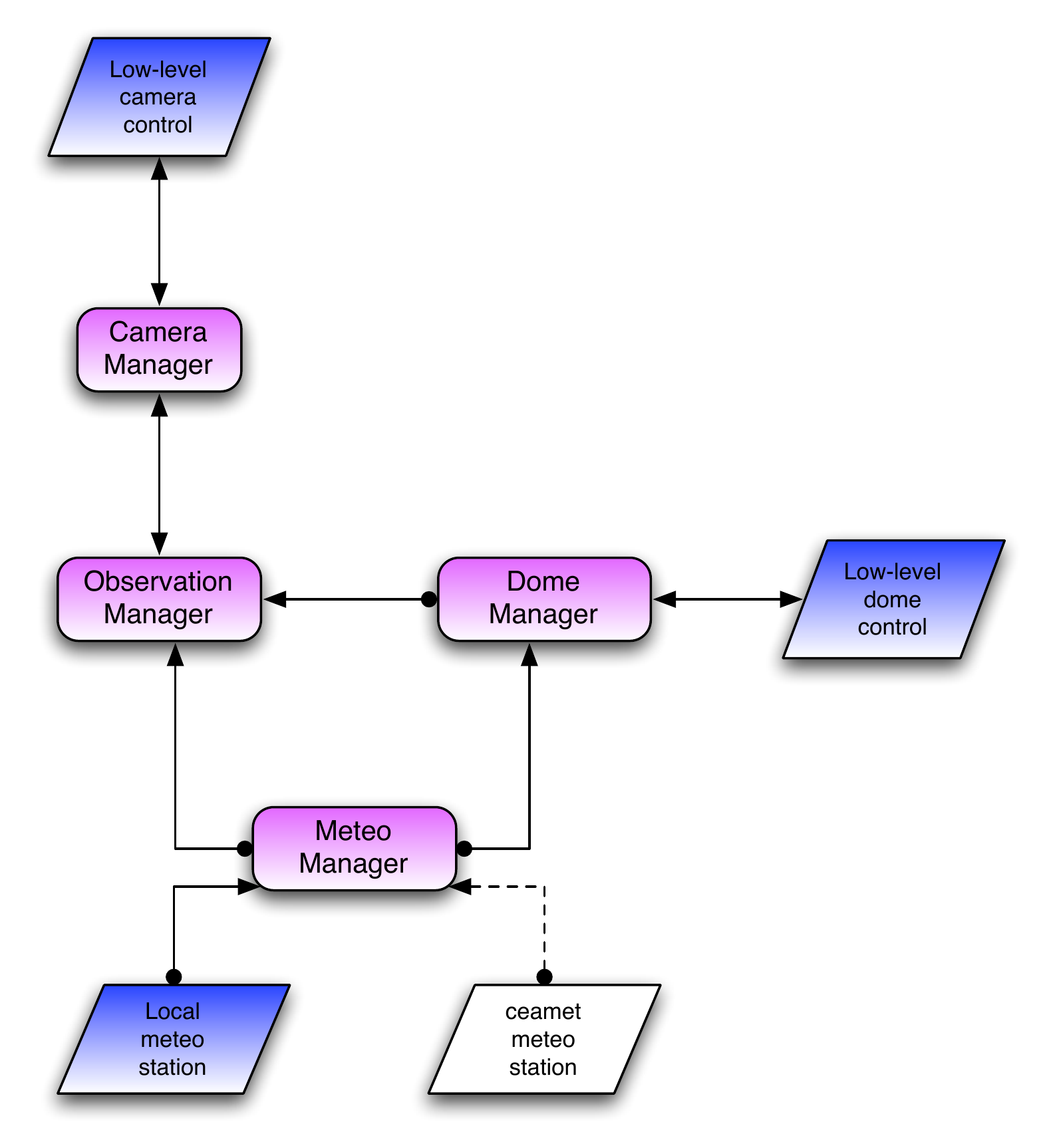}
%\picplace{5cm}{2cm} % Give the correct figure height and width in cm
%
\caption{Flux diagram of the main cycle of the observation manager.}
\label{fig:gen}       % Give a unique label
\end{figure}

The routine operations are managed by a set of 3 main programs, in communication one with the other through TCP/IP connection. The adoption of the TCP/IP paradigm allows a complete independence of the software from the running machine, hence allowing fast replacing of hardware in case of problems. These high level programs communicate with low-level processes managing the corresponding device (see fig. \ref{fig:gen}).

The organization into three main programs directly reflects what are the main duties an autonomous observatory should accomplish, namely check meteo conditions, safely manage the dome opening and, when all the conditions are good enough, send an observation to the telescope. In the following sections the realization of each one of these processes will be described.

\subsection{Meteo manager}
\label{sec:3}
Meteorological information is collected by a DAVIS Vantage Pro meteo station, installed on a pillar located 10m away from the dome; it is equipped with sensors for measuring air pressure, inner and outer air temperature and relative humidity, wind speed and direction, solar radiation and rain rate.

Meteo data is collected by a dedicated process, called \cmd{meteo\_man}. This program reads the meteo station every 5 minutes during the day and every minute during the night. Apart from reading data, it also checks if current meteo conditions are to be considered safe for operations; this requirement is achieved when all the following conditions are satisfied:
\begin{itemize}
	\item External humidity must be below 85\%
	\item Difference between the dew point and current external temperature as absolute value must be grater than 3 degrees Celsius
	\item Wind speed must not exceed 15 m/s. If the speed is between 12 m/s and 15 m/s then an azimuth limitation flag is raised, meaning that the telescope should avoid to point  directly into the wind and to the 180 degrees around it
	\item Atmosferic pressure must be above 870 hPa
	\item Rain rate must be equal to zero.
\end{itemize}

If any of the above conditions is not satisfied, then the \emph{bad meteo condition} flag is raised. Since the evaluation of safe meteo conditions is a very delicate task for any autonomous robotic telescope, particular prudence has been implemented in the algorithm evaluating the meteo conditions. Thus, when conditions are not satisfied, in order to consider meteo conditions good again, these must satisfy for 30 minutes the above rules with stronger constraints, namely humidity must stay below 80\%, wind speed below 12 m/s, pressure above 880 hPa and, of course, no rain.

\subsection{Dome manager}
\label{sec:dome}
The enclosure of the telescope is a classical hemispheric dome with two vertical sliding doors. The dome can rotate at a maximum speed of 6 degrees per second, which, during the pointing, translates to an average delay of 30 seconds respect to the telescope, so that the pointing of the telescope is poorly affected by the dome and it is not a concern at all for the scope of the scientific project.

The program \cmd{dome\_man}, which is in charge of managing the dome has two main duties:
\begin{itemize}
\item Provide a socket interface to the dome actions: although for robotical operations the dome is kept synchronized to the telescope position, so that it is not necessary to reposition the dome before each observation, its status is continuously monitored by another process, running on a different machine, which detects malfunctions and informs the team accordingly.
\item As autonomous process, it continuously acquires the meteo conditions and decides to open or close the dome depending on meteo conditions and sun altitude. Currently, the dome is allowed to stay opened when the sun in below -5 degrees of altitude.
\end{itemize}

\subsection{Observation manager}
\label{sec:5}
\begin{figure}[tb]
\centering
\includegraphics[height=12cm]{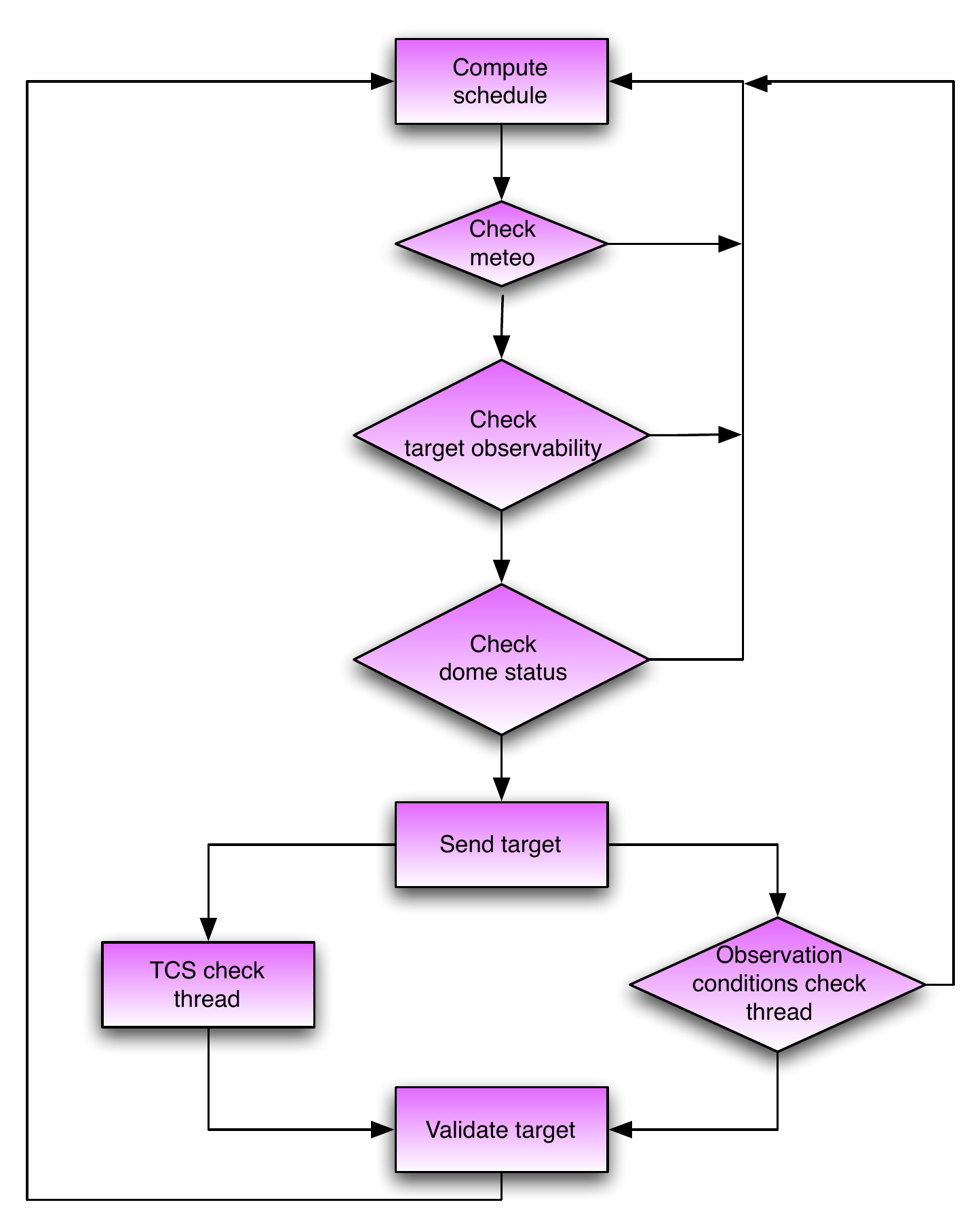}
%\picplace{5cm}{2cm} % Give the correct figure height and width in cm
%
\caption{Flux diagram of the main cycle of the observation manager.}
\label{fig:obs_man}       % Give a unique label
\end{figure}

The third main process, \cmd{obs\_man}, is the one which is in charge of managing the observations. Figure \ref{fig:obs_man} shows the diagram of the main activities. The system continuously checks if there is a target suitable for observations, if meteo conditions are fulfilled and if the dome is open. If this is the case, then the observation process starts. This is composed of two threads, one checking the tracking status of the telescope during all the exposure and the other checking that all the environmental conditions are valid during the whole observation. 

Due to a limitation in the low level software controlling the camera provided by the camera company, it is currently not possible to interrupt and ongoing observation. The information collected by the two threads is then used at the end of the exposure for validating the observation; if at least one of the two threads reported an error, then the target is ingested back into the archive for subsequent re-observation.

Although the scheduling software already takes into account constraints whose dissatisfaction can reveal dangerous for the telescope, the observation manager, before starting each target, performs a bunch of checks which prevent the telescope from directly pointing towards the sun, the moon and towards those directions which can not be reached by the telescope and would thus send the telescope to its limit switches.\\

 Each one of the above processes is coupled to an \cmd{isAlive} program. This is a crontab regulated job periodically checking that the corresponding process is running and, if this is not the case, re-starting it.

\subsubsection{Target archive and scheduling}

The set of targets is organized in a MySQL table. Each target can be associated to a user-defined priority value with the logic that the lowest is the value, the highest is the priority;  in addition the definition of all the most common constraints under which each target should be observed, such as the maximum moon fraction and distance, minimum and maximum airmass and periodicity of observation is implemented. 
\begin{figure}[tb]
\centering
\includegraphics[height=12cm]{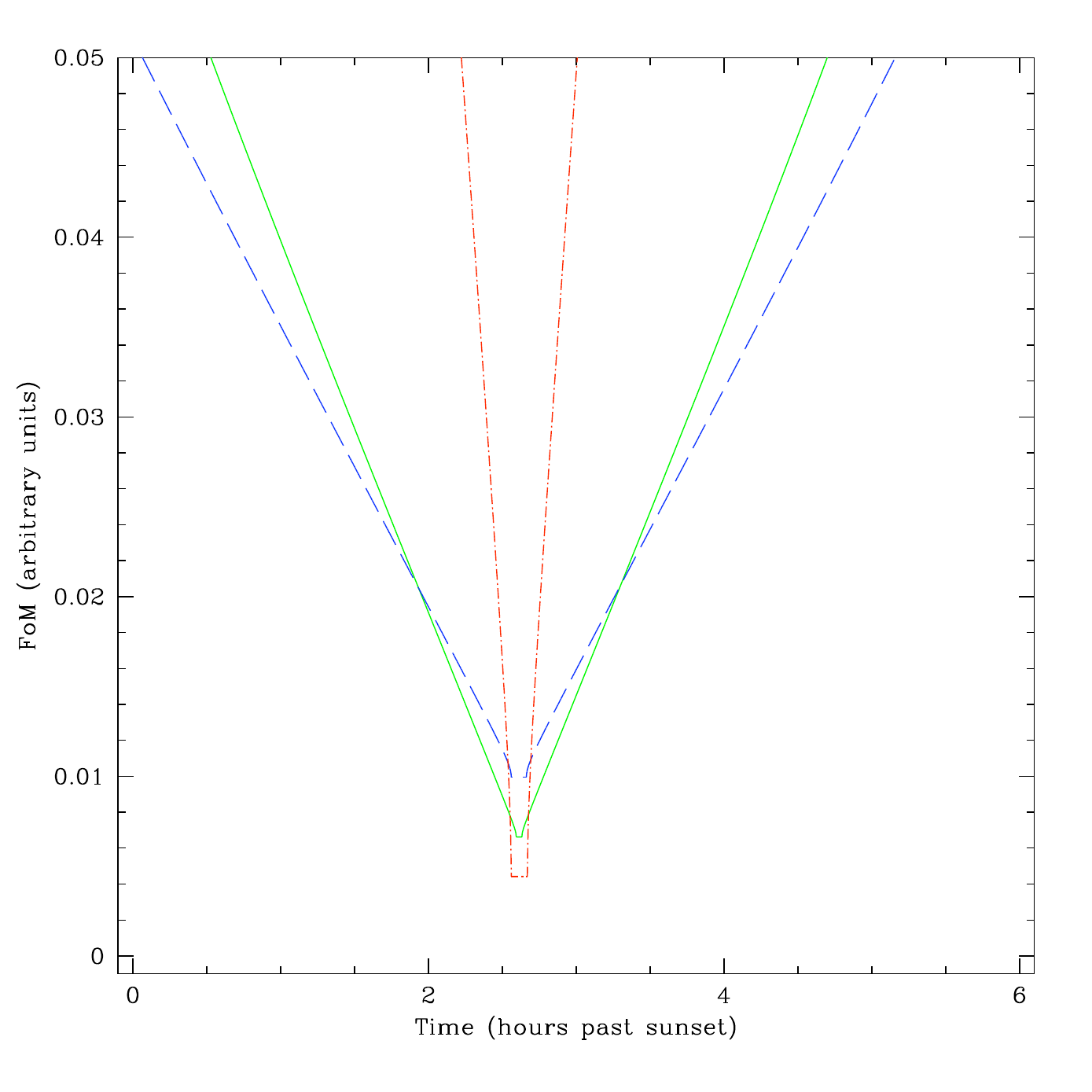}
%\picplace{5cm}{2cm} % Give the correct figure height and width in cm
%
\caption{Scheduler FoM for objects with the same priority=1, $RA=15h$ and $\delta=+10$, $+30$ and $+60$  for red dot-dashed, green solid and blue long dashed line respectively.}
\label{fig:sched}       % Give a unique label
\end{figure}

The fundamental idea behind the scheduling algorithm is to try to observe each target when it is closest to its minimum airmass (either because it is transiting through the meridian or because the user set a greater value in the minimum airmass field). This is achieved in two steps:
\begin{itemize}
	\item the scheduling software makes a first pass through all the targets, selecting only those whose constrains are satisfied from that moment and through all the exposure time;
	\item a figure of merit (FoM) is then applied to each of the selected targets . This FoM is a generalization of the $e^{-|x|}$ mathematical function, where $x$ here stands for the whole set of parameters. This is done in order to have a strong peak when all the conditions are best satisfied. The complete adopted function is:
	\begin{equation}
	h(p,f_a,f_t,f_d)=p (1-\exp(-(f_a/f_t)^{1/2})+f_d),
	\end{equation} 
	where $p \in{\mathcal{N}}$ is an integer number specifying the priority, with the value of $0$ denoting the highest priority; the normalized difference between the current and the minimum airmass $f_a$ is given by:
	\begin{equation}
	f_a=(a(0)-\min(a))/\min(a),
	\end{equation}
where $a(0)$ is the airmass at the time of the computation and $\min(a)$ is the maximum between the minimum airmass the object can reach on the sky and the user requirements for the minimum airmass; $f_t$ is the ratio between the total time an object is above its horizon (as define by the airmass constraints) and the required exposure time. Finally $f_d$ is the normalized distance of the object from the celestial pole. Explicitely, the expression for $f_d$ is:
\begin{equation}
f_d=n(\delta-\min(\delta))/(90.0-\min(\delta))
\end{equation}

Here $\delta$ is the declination of the object, $\min{\delta}$ is the minimum declination which is visible from the site for a given airmass and is $n$ a factor to reduce $f_d$ values to a suitable range of values of the whole $h$ function. 

The $f_d$ term has the effect of giving higher priority to those object with low declination,  which typically are the objects whose time for observation is shorter than that of objects at higher declinations.
	Finally, the $1/2$ exponent to the argument of the exponential function has been introduced in order to smooth the wings of the FoM.
	
	Figure \ref{fig:sched} shows the value of the result of applying the FoM described above to three targets with the same Right Ascension but with Declination $\delta=+10$, $+30$ and $+60$ respectively. 
	
	Note that targets with $p=0$ will always have the lowest value of $h$ (in fact $h\equiv0$ independently of the value of the other constraints), which translates into allocating the 0-priority target as soon as possible.\\
	
	The FoM expression can of course be generalized by introducing multiplicative parameters with the effect of shifting in time the points where the FoMs of targets with the same RA but different $\delta$ intersect (cfr. Figure \ref{fig:sched}), in order to better suite arising scheduling needs. This optimization will be done during the commissioning phase.
\end{itemize}

\subsection{Hardware}

All the software runs on almost \emph{of the shelf} hardware, without any particular need for high-speed processing, with a total number of three computers. The only exception is the machine with the \cmd{obs\_man} program which is running on a quad-core linux. This is due not because of the supposedly high demanding software of the observation manager, but because this machine is shared with another experiment which need real-time data analysis.

In addition, each machine has its own UPS system granting power supply to the computers in case of shortages. A UPS for the dome is also foreseen to be installed in the near future.

\section{Current status and future prospects}
The coding is now complete at the 80\% level and we foresee to complete it by the  end of the year. In particular one of the tasks we still need to implement is the procedure of automatic focusing. This will be achieved by a single exposure, tentatively at the beginning of the night, on which a bright star is imaged with different focus offset and by slightly moving the telescope between one acquisition and the next. The image will then be analyzed using SExtractor \cite{sex} and the fwhm of the star fitted by a parabola.

In parallel we started to develop a pipeline for the reduction of the survey data. This will make extensive use of the pyraf environment for the pre-reduction processes (bias subtraction, flat field correction etc.). Detection of the objects on each image will be managed by SExtractor; its mag\_auto value will be used as a first estimation of the flux, which will then be used by DoPhot \cite{dophot} to derive psf magnitudes.

\index{paragraph}
% Use the \index{} command to code your index words
\section{Conclusions}
\label{sec:5}
We described the software for the robotization of TROBAR, with the aim of performing a survey of the Galactic Plane, looking for H$\alpha$ emitting objects. The system is able to allocate and observe the targets in a MySQL archive when they are best visible and dependinfg on observing constraints. The code is now almost complete and we foresee to start robotic operations by the end of the present year.
\\
\newline
%{\it Acknowledgment.}

%
% 
% 
 
%
%
%
\end{document}